\documentclass{article}

\usepackage{arxiv}

\usepackage[utf8]{inputenc} 
\usepackage[T1]{fontenc}    
\usepackage{hyperref}       
\usepackage{url}            
\usepackage{booktabs}       
\usepackage{amsfonts}       
\usepackage{nicefrac}       
\usepackage{microtype}      
\usepackage{lipsum}		
\usepackage{graphicx}
\usepackage{natbib}
\usepackage{doi}
\usepackage{subfigure}

\title{Solid-amorphous transition is related to the waterlike anomalies in a fluid without liquid-liquid phase transition}

\author{ \href{https://orcid.org/0000-0002-8025-6529}{\includegraphics[scale=0.06]{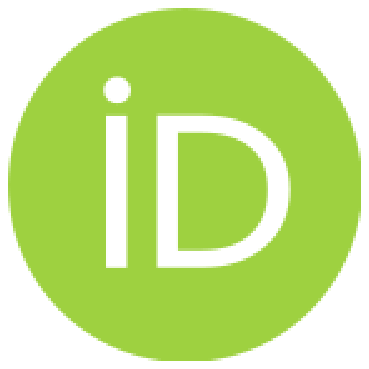} \hspace{1mm}José Rafael Bordin} \\
	Departamento de Física, Instituto de Física e Matemática,\\
	Universidade Federal de Pelotas\\
	Caixa Postal 354, CEP 96001-970, Pelotas, RS, Brazil \\
	\texttt{jrbordin@ufpel.edu.br} \\
	\And\href{https://orcid.org/0000-0000-0000-0000}{\includegraphics[scale=0.06]{orcid.eps}\hspace{1mm}\hspace{1mm}Leandro B. Krott} \\
	Centro de Ciências, Tecnologias e Saúde, Campus Araranguá\\
	Universidade Federal de Santa Catarina\\
	Rua Pedro João Pereira, 150, CEP 88905-120, Araranguá, SC, Brazil \\
	\texttt{leandro.krott@ufsc.br}
}

\date{\today}

\begin{document}
\maketitle

\begin{abstract}
The most accepted origin for the water anomalous behavior is the phase transition between two liquids (LLPT) in the supercooled regime connected to the glassy first order phase transition at lower temperatures. Two length scales potentials are an effective approach that have long being employed to understand the properties of fluids with waterlike anomalies and, more recently, the behavior of colloids and nanoparticles. These potentials can be parameterized to have distinct shapes, as a pure repulsive ramp, such as the model proposed by de Oliveira et al. [J. Chem. Phys. 124, 64901 (2006)]. This model has waterlike anomalies despite the absence of LLPT. To unravel how the waterlike anomalies are connected to the solid phases we employ Molecular Dynamics simulations. We have analyzed the fluid-solid transition under cooling, with two solid crystalline phases, BCC and HCP, and two amorphous regions being observed. We show how the competition between the scales creates an amorphous cluster in the BCC crystal that leads to the amorphization at low temperatures. A similar mechanism is found in the fluid phase, with the system changing from a BCC-like to an amorphous-like structure in the point where a maxima in $k_T$ is observed. With this, we can relate the competition between two fluid structures with the amorphous clusterization in the BCC phase.Those findings help to understand the origins of waterlike behavior in systems without liquid-liquid critical point.

\end{abstract}

\section{\label{sec:level1}Introduction}

Every known fluid solidifies upon cooling. However, the cooling process can lead to solid structures with distinct types of ordering. While crystalline solid phases show a long-range, macroscopic order, in condensed amorphous phases, as a glass or amorphous solid, no long-range order is observed~\cite{Debenedetti97,icebook05}. In fact, understanding the glass transition and the behavior of fluids in the vicinity is a huge task in condensed matter physics; and the challenge grows for polyamorphic systems. A one-component system has polyamorphism if it has the coexistence (often metastable) of two amorphous phases~\cite{kivelson02,braz03, loer09, Angell14,anisimov18,sander18, Bachler19,McMillan21,Hernandes22}.

Certainly, the most famous polyamorphous material is water. There are growing evidences that, deep into the supercooled liquid region, water have a liquid–liquid first-order phase transition that ends in a liquid-liquid critical point (LLCP), which is connected to the first order phase transition observed between two glassy phases at lower temperatures~\cite{Mishima85, poole1992, Stanley2009, Giovambattista12, gallo2016, kim20, shi20,Bachler21, Foffi21,giovambattista21,Caupin21,lucas22,Verde2022}. 
It is possible to investigate the location of a critical point through an analysis of the thermodynamic response functions. For instance, the specific heat and isothermal compressibility display a line of maxima in the P-T plane (Widom's line) that typically ends at the critical point where the maximum evolves into a divergence in the thermodynamic limit~\cite{simeoni10,Brazhkin11, brazhkin18,zeron19, Losey19, bianco19}. Water, in addition to the Widom line ending at the vapor-liquid critical point, shows a second ending at the LLCP~\cite{Xu2005, franzese07,stanley2008,Kumar2008,abascal10, luo2015,gallo2016}. Predicted in 1992 by Poole and co-authors~\cite{poole1992}, experimental studies to observe the LLCP have been hampered by the spontaneous crystallization in the no-man's land for 25 years. Only recently the Widom line was detected in experiments~\cite{kim2017,kim20,winker23, kringle20}. Nevertheless, many other systems have evidences of polyamorphism, as silicon~\cite{sastry03, beye10,yagyik21}, silica~\cite{saika04,lascaris14,chen17, Trang20,Yu22}, hydrogen~\cite{zaghoo16, nathaniel22}, sulfur~\cite{plas15, henry20,shumov22}, phosphorous~\cite{yoshi00, yoshi04,yang21}, carbon~\cite{james99, sun21}, triphenyl phosphite~\cite{tanaka04}, gallium~\cite{liu21}, molten elements~\cite{braz99}, metallic glasses~\cite{sheng07,greaves08, li22}, germania~\cite{smith95}, tellurium~\cite{sun22} and more~\cite{angell95,rizzatti19,tanaka20,lucas21,desgranges18}.

The existence of LLCP is the mainly accepted scenario to explain the anomalous behavior of water~\cite{gallo21}. In fact, water is the most anomalous known material, with more than 70 thermodynamic, dynamic and structural behaviors that strongly diverge from the observed in most fluids~\cite{url,rudolf2011}.  For instance, it is long known that water  density increases as the temperature grows from 0$^o$C to 4$^o$C at 1 atm ~\cite{Ke75}, whereas in most materials heating is associated naturally with the thermal expansion. Once most of these anomalies, and the LLCP, occur in the supercooled regime, and due  to the difficulty to access this region experimentaly~\cite{gallo21}, computer simulations have long been applied to overcome the experimental barriers and understand the behavior of water and others anomalous fluids~\cite{Stanley2009,gallo2016,Liu12,gallo21}. In this sense, many models are able to reproduce the water polyamorphism and the LLCP~\cite{giovambattista21, Palmer13a,Palmer18b,chiu13}. It is interesting that models without LLCP, such as the  SPC/E, show a low density amorphous (LDA) - high density amorphous (HDA) phase transition and waterlike anomalies~\cite{Giovambattista12}.

Water molecules interaction are known to be computationally expensive and difficult to treat, especially in many body atomistic models~\cite{abella23}. Is this way, many isotropic core-softened potentials with two characteristic length scales have been employed to unveil the waterlike behavior, searching for the origins of anomalies~\cite{franzese2011}. Since the early 1970's, where Hemmer and Stell~\cite{hemmer1970} observed that we can induce polyamorphism by softening a hard core interaction judiciously, obtaining a core-softened (CS) potential, passing by the seminal works by Jagla in the late 1990's~\cite{Ja98,jagla1999b,jagla1999b}, many spherical symmetric core-softened potentials have been employed to study polyamorphous systems~\cite{xu10,gnan14,xu19,dasgupta20,fomim20} and to understand the relation between waterlike behavior and the competition between distinct structures in the fluid~\cite{yan06,gibson2006,franzese2007,Xu11,reisman13,das13,krott15,luo2015, bordin15, Bordin2016b, Leo17,haro18,saki18, Ryzhov20, roca22,bretonnet22}. In fact, these simple models can be applied for a variety of systems~\cite{likos01}. Experimental works have shown that the effective interaction in solutions of pure or grafted PEG colloids are well described by core-softened potentials~\cite{colloid1, colloid2, Haddadi20}, and computational studies indicate the same type of effective interactions for polymer-grafted nanoparticles~\cite{marques2020a, Lafitte2014} or star polymers~\cite{Bos19}.  
 
 The existence of two length scales in the CS potential does not guarantee the existence of waterlike anomalies~\cite{quigley05}. The competition between the length scales is a key ingredient to observe the anomalous thermodynamic, dynamic and structural behavior in such fluids~\cite{alan2008,franzese2007,Stanley2009,jonathas, ney09,Marques2020, Marques21a}, reproducing the competition between two liquids in supercooled water~\cite{gallo2016, gallo21}. However, for purely repulsive potentials, as the one proposed by de Oliveira and co-authors, no LLCP was observed - despite the competition between the scales and the waterlike anomalies~\cite{alan2008,alan2006,jonathas}. This is a similar scenario from the observed in SPC/E water~\cite{Giovambattista12} - however, the existence of solid poly(a)morphism and their relation with the waterlike anomalies were not investigated. 

 To unravel the solid phases and the relation with the waterlike anomalous behavior, we perform large scale $NpT$ and $NVT$ Molecular Dynamics simulations cooling a purely repulsive CS from the fluid to the solid phase. To this end, we employ the model proposed by de Oliveira et al.~\cite{alan2006}, reproducing the waterlike anomalies obtained in previous works~\cite{alan2006, BarrosDeOliveira2006, DeOliveira2006, alan2008, jonathas}. We expand these works by looking to the low temperature regime.  Four distinct solids structures were observed: two crystalline and two amorphous. Our results indicate that the system undergoes a fluid-crystal solid first order phase transition, and a second order fluid-amorphous transition. Also, we show how the anomalies are connected to a crystal-amorphous phase transition, and found a maxima in the isothermal comprehensibility that does not end in a LLCP but it is connected to the amorphization of a crystal structure. 

 The remainder of the paper is organized as follows. In Section II we present our anomalous fluid model and summarize the details of the simulations. Next, in section III our most significant results. In particular, we will focus on the solidification scenario for the distinct solid structures and in their relations with waterlike anomalies. The paper is closed with a brief summary of our main conclusions and perspectives.

\section{Model and Simulation Details}\label{Model}
\subsection{The Model}

\begin{figure}[h!]
    \centering
    \includegraphics[width=0.40\textwidth]{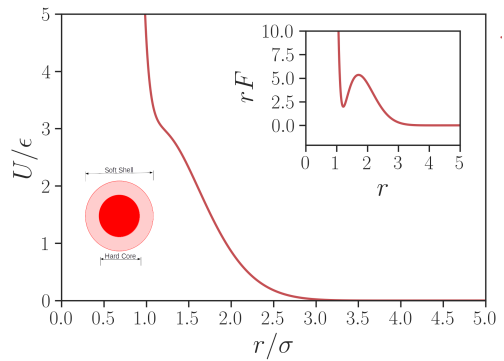}
    \caption{Ramp with shoulder potential. Graph inset: the force ($F = -\partial U/\partial r$) times the distance $r$. Molecule insets is the schematic depiction of the hard core-soft shell particle.}
    \label{fig:potential}
\end{figure}

Once it is an effective model, and for the sake of simplicity, all quantities will be depicted in standard LJ units~\cite{AllenBook}. The fluid particles interaction was modeled by the CS potential given by~\cite{BarrosDeOliveira2006,DeOliveira2006}  
     $$
     U_(r) = 4\epsilon \bigg[\bigg(\frac{\sigma}{r}\bigg)^{12} - \bigg(\frac{\sigma}{r}\bigg)^{6}  \bigg] +
     $$
        \begin{equation}
                u_0 \exp\bigg[-\frac{1}{c_{0}^{2}}\bigg(\frac{r-r_0}{\sigma}\bigg)^2\bigg],
        \label{eq:CS}
\end{equation}

\noindent which is composed of a short-range attractive Lennard-Jones potential and a repulsive Gaussian potential centered at $r_0$, with depth $u_0$ and width $c_0$. Using the parameters $u_0 = 5\epsilon$, $c_0^2 = 1.0$, and $r_0/\sigma = 0.7$~\cite{BarrosDeOliveira2006,DeOliveira2006,Bordin2018} the potential (\ref{eq:CS}) acquire the purely repulsive ramp-like shape shown in Fig.\ref{fig:potential}. The waterlike anomalous behavior in this potential arises from the competition between two characteristic length scales~\cite{jonathas,Bordin2018}: a closer one located at the shoulder in $r_1 = 1.2\sigma$, where the force has a local minimum (as we can see in Fig.~\ref{fig:potential} graph inset), while the further length scale is located at $r = 2.0\sigma$, where the fraction of imaginary modes of the instantaneous normal modes spectrum has a local minimum\cite{BarrosDeOliveira2010}. The cutoff radius for the interaction is $r_c = 3.5\sigma$.

\subsection{Simulation Details}

The phase diagram was obtained by simulations in the isothermal-isobaric ensemble ($NpT$) with fixed number of particles $N= 5000$, as implemented in the ESPResSo package~\cite{arnold2006,arnold2013}. For the $NpT$ simulations, isobars raging from $p = 0.0001$ up to $p = 4.00$ with distinct intervals were simulated. Along each isobar we perform a cooling process, starting at $T = 0.600$, reducing the temperature up to $T = 0.025$ with a temperature step $\delta T = 0.025$. A random system is created at the highest temperature - at this temperature, the system is always in the fluid state~\cite{BarrosDeOliveira2006}. Then,  $5 \times 10^6$ time steps in the $NVT$ ensemble thermalizes the system, followed by $1 \times 10^6$ time steps in the $NpT$ ensemble to equilibrate the system's pressure and density. $1 \times 10^7$ time steps are then performed for the production of the results, with averages and snapshots being taken at every $1 \times 10^5$ steps. After that, we cool down the system to $T - \delta T$ along the isobar, repeating the thermalization and the equilibration process in the $NVT$ and $NpT$ ensembles, respectively, in order to relax the system at the new temperature and achieve the proper new density before run the production steps. In this way, the two equilibration steps allow the system to cool down and relax at the new state. To ensure that the system is in the equilibrium we averaged quantities as energy, entalpy, density, instantaneous pressure and kinetic temperature along the simulations, including in the equilibration processes.  The Langevin thermostat~\cite{allen2017}, that keeps the temperature fixed in our ESPResSo implementation, has a coupling parameter $\gamma_0=1.0$. The piston parameters for the Anderson barostat~\cite{allen2017} are $\gamma_p=0.0002$ and mass $m_p = 0.001$. Here, the molecule density $\rho$ is defined as $N/V$ with $V$ being the mean volume at a given pressure and temperature.

To confirm the low temperature phase transitions, simulations using the isothermal ensemble ($NVT$) were performed in the range $0.020 \le T \le 0.250$ and $0.048 \le \rho \le 0.296$ as implemented in a fortran90 code developed in the house~\cite{Krott2013, Krott2015}. The temperature was controlled using the Nos\`e-Hoover thermostat~\cite{Frenkel} with coupling parameter $Q = 2.0$. Two different initial configurations of the systems were simulated: solid and liquid states. Using different initial configurations allow us to identify precisely the final state of the system, avoiding metastability. The equilibrium state was reached after $4\times10^6$ steps, followed by $2\times10^6$ simulation run. We used a time step $\delta t$ = 0.001. All the physical quantities were obtained from 50 uncorrelated samples. To check the stability of the systems, we verified the energy as a function of time and the pressure, evaluated by the virial, as a function of density.

The temperature of maximum density (TMD) along each isobar was used to characterize the density anomaly. Also, the locus of the maximum of response functions were obtained: the isothermal compressibility $\kappa_T$, the isobaric expansion coefficient $\alpha_P$ and the specific heat at constant pressure $C_p$ and at constant volume $C_V$,

$$
    \kappa_T= \frac{1}{\rho} \left ( \frac{\partial \rho}{\partial p} \right )_T\;,  \quad 
    \alpha_p= -\frac{1}{\rho} \left ( \frac{\partial \rho}{\partial T} \right )_p\;,
    $$
    \begin{equation}
    C_p = \frac{1}{N} \left(\frac{\partial H}{\partial T} \right)_p\;, \quad C_V = \frac{1}{N} \left(\frac{\partial U}{\partial T} \right)_V\;.
\end{equation}

\noindent Here,  $H = U + PV$ is the system entalpy. The quantities shown  were obtained by numerical differentiation and,  as consistency check, using statistical fluctuations: the compressibility as the measure of volume fluctuations, the isobaric heat capacity is proportional to the entropy fluctuations experienced by $N$ particles at fixed pressure, and the thermal expansion coefficient reflects the correlations between entropy and volume fluctuations ~\cite{allen2017,tuckerman2010}. 

In order to describe the connection between structure and thermodynamics, we have analyzed translational and orientational properties. The radial distribution function (RDF) $g(r)$ was evaluated and subsequently used to compute the excess entropy $s_{ex}$ and the translational order parameter $\tau$. The first one can be obtained by counting all accessible configurations for a real fluid and comparing it with the ideal gas entropy \cite{dzugutov1996}. Then, $s_{ex}$ is a negative quantity that increases  with  temperature, as the  full entropy $S$ does  \cite{dyre2018,dyre2020}.  Expanding  $s_{ex}$ in terms of two-particle, three-particle contributions, etc.~\cite{galliero2011}, we have
\begin{equation}
	s_{ex} = s_2 + s_3 + s_4+... 
\end{equation}

The two-particle contribution, 
\begin{equation}
s_2=-2\pi \rho \int_{0}^{\infty}\left [ g(r)\ln g(r) - g(r)+1) \right ]r^2dr.
\end{equation} 
corresponds to $85\%$ and $95\%$ of the total excess entropy in Lennard-Jones systems, being the dominant contribution to excess entropy \cite{raveche1971,baranyai1989,sharma2006}. 

In order to ascertain how the short and long range coordination shells behaves, we evaluated the cumulative two-body entropy~\cite{Krek08,Cardoso21} 
\begin{equation}
\label{cs2}
C_{s_2}(r) = -2\pi\rho \int_0^r [g(r')\ln(g(r')) - g(r') +1]r'^2 dr'\;.
\end{equation}
Here, $r$ is the upper integration limit. It is relevant to address that the two-body excess entropy is not a thermodynamic quantity, but a structural order metric that connects thermodynamics and structure. Then, $C_{s_2}$ is a useful tool to analyze long-range structural characteristics of the core-softened system.

The translational order parameter is defined as~\cite{Er01}
\begin{equation}
\label{order_parameter}
\tau \equiv \int^{\xi_c}_0  \mid g(\xi)-1  \mid d\xi,
\end{equation}
\noindent where $\xi = r\rho^{1/3}$ is the interparticle distance $r$ scaled with the average separation between pairs of particles $\rho^{1/3}$. $\xi_c$ is a cutoff distance, defined as $\xi_c = (L\rho^{1/3})/3$, where $L$ is the simulation box size. For an ideal gas (completely uncorrelated fluid), $g(\xi) = 1$ and $\tau$ vanishes. For crystals or fluids with long range correlations $g(\xi) \neq 1$ over long distances, which leads to $\tau >0$. The excess entropy and the translational order parameter $\tau$ are linked, once both are dependent on the deviation of $g(r)$ from unity. The translational structural anomaly can then be characterized by an anomalous decrease of $\tau$ under compression.

Another structural quantity evaluated was the orientational order parameter (OOP), that gives insight on the local orientational order~\cite{steinhardt1983, Er01, DeOliveira2006, Yan05}. The OOP for a specific particle $i$ with $N_b$ neighbors is given by
\begin{equation}
\label{OOP_ql}
q_{l} (i) = \sqrt{\frac{4 \pi}{2l + 1} \sum_{m = -l}^{l} \vert q_{lm} \vert^2 },
\end{equation}
with
\begin{equation}
\label{OOP_qlm}
q_{lm} (i) = \sqrt{\frac{1}{N_{b}} \sum_{j=1}^{N_{b}} Y_{lm}(\theta(\vec{r}_{ij}),\phi(\vec{r}_{ij})) }.
\end{equation}
where $Y_{lm}$ are the spherical harmonics of order $l$ and $\vec{r}_{ij}$ is the distance between two particles. The OOP for a whole system is obtained by taking the average over the parameter value for each particle $i$, $\tilde{q}_l = \langle q_l (i) \rangle _i $. In this work we evaluated the OOP for $l = 4$ and 6, using the Freud python library \cite{freud2020} -- the number of neighbors for each particle was found computing Voronoi diagrams using voro++ \cite{voro++}, as explained in details in our previous work~\cite{Hernandes22}. The anomalous behavior is characterized by a maxima in $\tilde{q}_6$ and a minima in $\tilde{q}_4$

The dynamic behaviour was analyzed by the mean square displacement (MSD), given by
\begin{equation}
\label{r2}
\langle [\vec r(t) - \vec r(t_0)]^2 \rangle =\langle 
\Delta \vec r(t)^2 \rangle\;,
\end{equation}
where $\vec r(t_0) =$ and  $\vec r(t)$ denote the particle position at a time $t_0$ and at a later time $t$, respectively. The MSD is then related to the diffusion coefficient $D$ by the Einstein relation,
\begin{equation}
 D = \lim_{t \rightarrow \infty} \frac{\langle \Delta 
\vec r(t)^2 \rangle}{6t}\;.
\end{equation}
\noindent The diffusion anomaly region was obtained using the $(D, p)$ isotherms. For regular fluids $D$ decreases by compressing. However, for a range of pressures, $D$ increases with $p$ in fluids with diffusion anomaly. With this, the diffusion anomaly extrema (DE) may be defined using the minima and maxima in these curves~\cite{alan2008,franzeseJCP2010,marques2020a}.

Finally, the onset of crystallization was monitored analyzing the local structural environment of particles  by means of the Polyhedral Template Matching (PTM) method implemented in the Ovito software~\cite{Larsen2016,ovito}. 

\section{Results and Discussion}
\label{Results}

\subsection{The Fluid-Solid Transitions}

\begin{figure}[h!]
    \centering
    \includegraphics[width=0.475\textwidth]{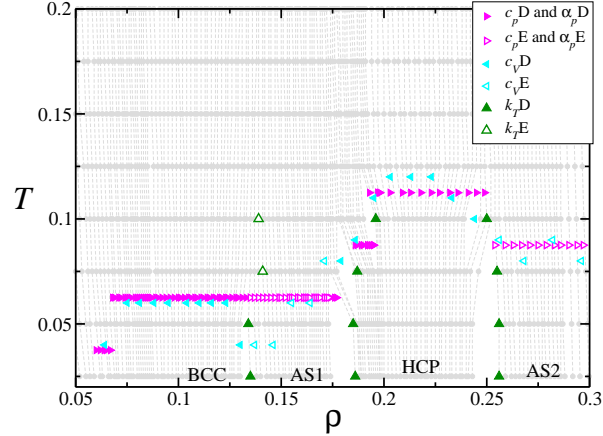}
    \caption{Low temperature $T\times\rho$ phase diagram obtained from the $NpT$ simulations. Isobars ranging from $p = 0.050$ up to $p = 4.200$ are shown in dashed grey lines with circular points - the deviations from the mean value of $\rho$ are smaller than the data points. The Extrema (E) -- that can be a maxima or a minima -- and the Discontinuity (D) observed in the response functions are indicated by the points: magenta filled right triangle for a discontinuous behavior in the $c_p\times T$ and $\alpha_p\times T$ isobars, magenta open right triangle for a extrema in the $c_p\times T$ and $\alpha_p\times T$ isobars, cyan filled left triangle for a discontinuous behavior in the $c_V\times T$ isochores, cyan open left triangle for a maxima in the $c_V\times T$ isochores, green filled up triangle for a discontinuous behavior in the $k_T\times p$ isotherm, green open up triangle for a maxima in the $k_T\times p$ isotherms.}
    \label{fig:lowTPD}
\end{figure}

\begin{figure}[h!]
    \centering
    \includegraphics[width=0.475\textwidth]{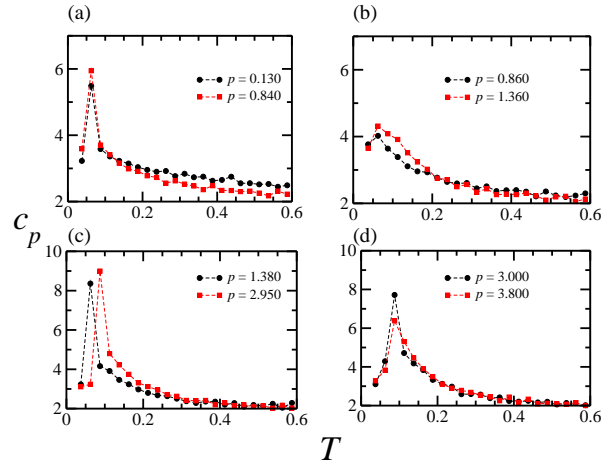}
    \caption{Dependence of $c_p$ as function of $T$ for two distinct isobars that crosses the (a) fluid-BCC solid discontinuous phase transition, (b) fluid-amorphous solid 1 continuous phase transition, (c) fluid-HCP solid discontinuous phase transition, and (d) fluid-amorphous solid 2 continuous phase transition.}
    \label{fig:diagramcp}
\end{figure}

 Although anomalous, fluid polymorphism was not reported for the  purely repulsive CS model employed in this work~\cite{DeOliveira2006, BarrosDeOliveira2006, jonathas}.  One possibility is that the fluid properties vary very smoothly, with the fluid near a crystalline phase behaving distinctly from water near an amorphous phase.  In the original works by de Oliveira and co-authors~\cite{DeOliveira2006, BarrosDeOliveira2006}, and in their next articles~\cite{alan2008a, jonathas}, the crystallization limit was not explored. Recently, Bordin and co-authors~\cite{Cardoso21,Nogueira22} explored the 2D limit, and found three distinct crystalline solid phases separated by two amorphous solid phases. To check if a poly(a)morphism will be observed in the 3D limit, we can first look to the specific heat behavior along an isobar (for the $NpT$ simulations), and corroborate the findings evaluation the specific heat at constant volume using the $NVT$ simulations. In this way, we show in Fig.~\ref{fig:lowTPD} the $T\times \rho$ phase diagram for temperatures below $T = 0.250$. Here, the gray circles connected by gray dashed lines indicate the behavior of the isobar upon cooling and, to identify the fluid-solid phase transitions, the response functions were evaluated. Interesting, both $c_p$ and $\alpha_p$ can show discontinuities ($c_pD$ and $\alpha_pD$) or extrema ($c_pE$ and $\alpha_pE$) depending on the isobar - in a similar way, $c_v$ can also have extrema or discontinuities along distinct isochores.
 
\begin{figure}[h!]
    \centering
    \includegraphics[width=0.475\textwidth]{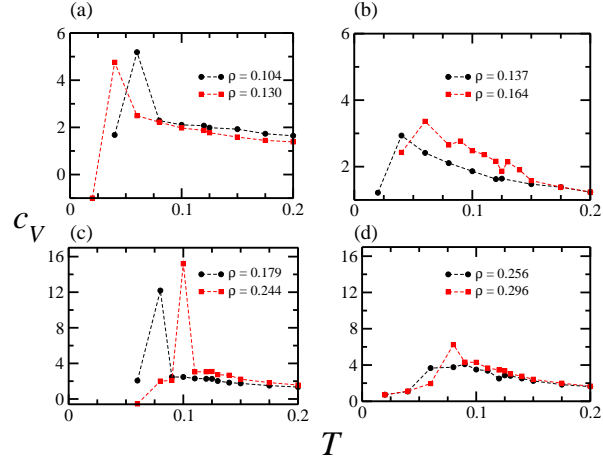}
    \caption{Dependence of $c_V$ as function of $T$ for two distinct isochores that crosses the (a) fluid-BCC solid discontinuous phase transition, (b) fluid-amorphous solid 1 continuous phase transition, (c) fluid-HCP solid discontinuous phase transition, and (d) fluid-amorphous solid 2 continuous phase transition.}
    \label{fig:diagramcv}
\end{figure}

\begin{figure}[h!]
    \centering
    \includegraphics[width=0.475\textwidth]{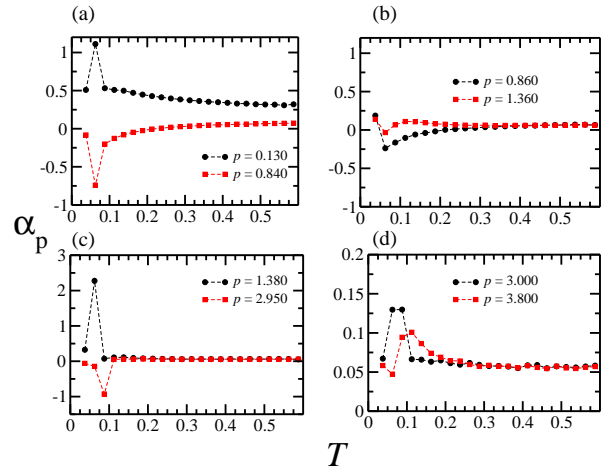}
    \caption{Dependence of $\alpha_p$ as function of $T$ for two distinct isobars that cross the (a) fluid-BCC solid discontinuous phase transition, (b) fluid-amorphous solid 1 continuous phase transition, (c) fluid-HCP solid discontinuous phase transition, and (d) fluid-amorphous solid 2 continuous phase transition.}
    \label{fig:diagramap}
\end{figure}

In Fig.~\ref{fig:diagramcp} we show the $c_p$ behavior. At low pressures, up to $p = 0.840$, the specific heat has a discontinuity (D) at low temperatures, indicating a first order phase transition, as we can see in Fig.~\ref{fig:diagramcp}(a). Analysing isochores in the same region, as illustrated in Fig.~\ref{fig:diagramcv}(a), we can see that $c_V$ also has a discontinuity in a close temperature. On the other hand, for pressures ranging from $p = 0.860$ to 1.360, the $c_p$ has a maxima. This $\lambda-$like curve is typical from second order phase transitions, and was observed for both $c_p$, Fig.~\ref{fig:diagramcp}(b), and $c_v$, Fig.~\ref{fig:diagramcv}(b). Compressing the system, the discontinuous behavior is recovered for both specific heats, see Fig.~\ref{fig:diagramcp}(c) and Fig.~\ref{fig:diagramcv}(c) and, for the higher pressures and densities, it changes once again to a $\lambda$ shape. Such behavior was also observed in another response function, $\alpha_p$, shown in Fig.~\ref{fig:diagramap}. Here is also possible to observe two waterlike features of this model: the density anomaly and a fluid phase denser than the solid phase (as the isobars in Fig.~\ref{fig:lowTPD} shown). 

\begin{figure}[h!]
    \centering
     \includegraphics[width=0.475\textwidth]{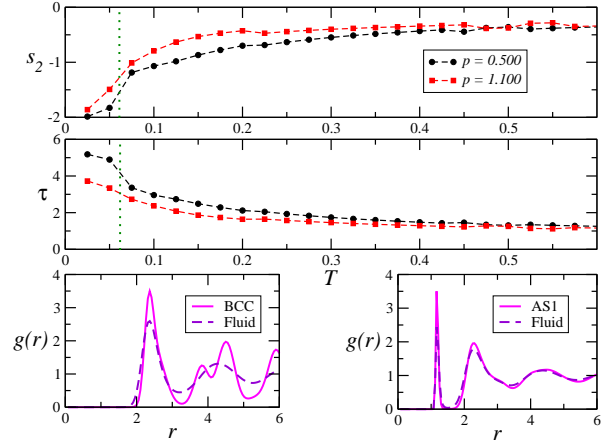}
    \caption{Behavior of $s_2$ (upper panel) and $\tau$ (middle panel) as function of temperature for the isobars $p=0.500$ and $p=1.100$. Lower panel: at left we show the RDF for the system in the BCC phase in the point $p=0.500, T=0.050$ and in the fluid phase at $p=0.500, T=0.075$, and in the right the RDF for the system in the amorphous solid phase 1 in the point $p=1.100, T=0.050$ and in the fluid phase at $p=1.100, T=0.075$.}
       \label{fig:s2tauisobarslow}
\end{figure}

The discontinuous/continuous transitions can also be observed using structural parameters. First, we analyze the behavior of $s_2$ and $\tau$ along two pressures isobars. The first one, $p = 0.500$, has a discontinuity in the response functions $c_p$ and $\alpha_p$, and the second, $p = 1.100$, a continuous behavior. By looking to the $s_2$ and $\tau$ curves, shown in Fig.~\ref{fig:s2tauisobarslow}, we can also see a discontinuous behavior along the fluid-solid transition, evidenced by the abrupt change in the RDF behavior in the lower panel of Fig.~\ref{fig:s2tauisobarslow}. Cooling down the system from $T = 0.075$ to $T = 0.050$ leads to an abrupt change in the RDF, from a fluid to a solid with defined structure -- with the PTM method~\cite{Larsen2016,ovito} and $q_6$ and $q_4$ analyzes~\cite{lechner2008} we identified as a body-centered cubic (BCC) structure.  On the other hand, along the isobar $p=1.100$ the translational structural parameters $s_2$ and $\tau$ change continuously once the structure at the amorphous phase is similar to the fluid structure, as shown in the RDF of the lower panel of Fig.~\ref{fig:s2tauisobarslow}. This amorphous region will be denominated as AS1.

\begin{figure}[h!]
    \centering
\includegraphics[width=0.475\textwidth]{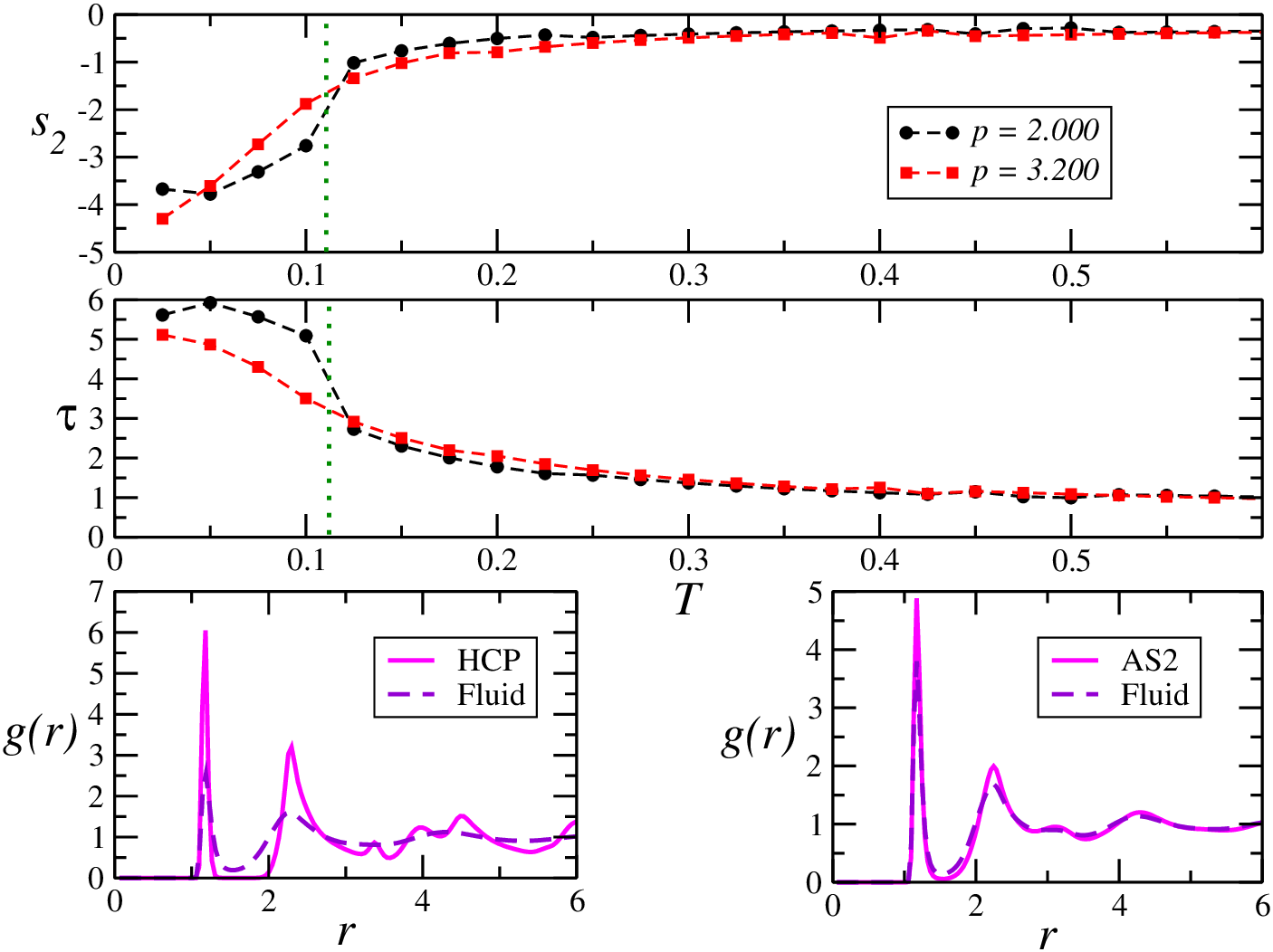}
    \caption{Behavior of $s_2$ and $\tau$ ad function of temperature for the isobars $p=2.000$ and $p=3.200$. Lower panel: at left we show the RDF for the system in the HCP phase in the point $p=2.000, T=0.100$ and in the fluid phase at $p=2.000, T=0.125$, and in the right the RDF for the system in the amorphous solid phase 2 in the point $p=3.200, T=0.075$ and in the fluid phase at $p=3.200, T=0.100$. }
       \label{fig:s2tauisobarshigh}
\end{figure}

Compressing the system, $s_2$ and $\tau$ show a discontinuous behavior along the isobar $p = 2.000$ and continuous at $p=3.200$ in Fig.~\ref{fig:s2tauisobarshigh}, as the thermodynamic response functions. The discontinuous behavior can be associated to a fluid-solid transition, with a hexagonal closed packed (HCP)~\cite{lechner2008, Larsen2016,ovito} structure in the solid crystalline phase, while the continuous behavior correspond to a fluid-amorphous transition - the AS2 phase. As in the fluid-AS1 transition, the main difference between the fluid and AS2 structures is the occupancy in the two characteristic length scales, as the RDF in the lower panel of  Fig.~\ref{fig:s2tauisobarshigh} show: the amorphous phases have higher peaks at $r_1$ and $r_2$ and the occupancy between the peaks vanishes, indicating the solidification.

\begin{figure}[h!]

    \centering
    \subfigure[]{\includegraphics[width=0.475\textwidth]{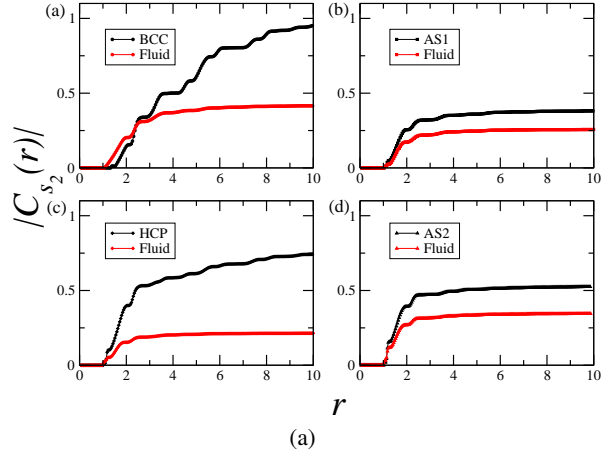}}
    \caption{Cumulative two-body entropy in the vicinity of the phase transitions. In (a) we show $|C_{s2}(r)|$ for the point $p=0.500, T=0.050$ in the BCC phase and $p=0.500, T=0.075$ in the fluid phase, (b) for $p=1.100, T=0.050$ in the AS1 phase and $p=1.100, T=0.075$ in fluid phase, (c) HCP phase in the point $p=2.000, T=0.100$ and fluid phase at $p=2.000, T=0.125$, and (d) AS2 phase at $p=3.200, T=0.075$ and fluid phase at $p=3.200, T=0.100$.}
       \label{fig:cums2isobar}
\end{figure}

The existence of two crystalline solid phases and two amorphous phases can also be inferred from the longer range translational ordering. For a ordered solid, $|C_{s2}(r)|$ increases with $r$, while for disordered systems - as fluid or amorphous solids -- $|C_{s2}(r)|$ reaches a plateau as the RDF$\rightarrow$ 1.0 at short distances. It can be seen in the cumulative two-body entropy, Fig.~\ref{fig:cums2isobar}: the BCC and HCP phases entropy continually increase until large distances $r$, indicating a longer range ordering in the coordination shells. On the other hand, the fluid and amorphous solid phases reach a plateau after a short distance, characterizing a shorter range ordering. 

\begin{figure}[h!]
    \centering
     \includegraphics[width=0.475\textwidth]{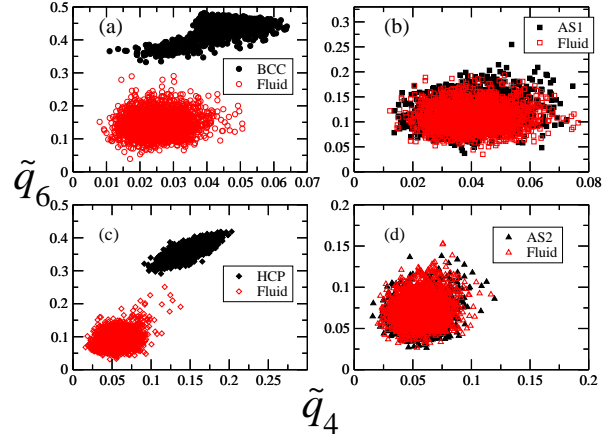}
        \caption{$\tilde q_6 \times \tilde q_4$ for (a) the point $p=0.500, T=0.050$ in the BCC phase and $p=0.500, T=0.075$ in the fluid phase, (b) for $p=1.100, T=0.050$ in the AS1 phase and $p=1.100, T=0.075$ in fluid phase, (c) HCP phase in the point $p=2.000, T=0.100$ and fluid phase at $p=2.000, T=0.125$, and (d) AS2 phase at $p=3.200, T=0.075$ and fluid phase at $p=3.200, T=0.100$. Each point corresponds to one particle in the system.}
       \label{fig:q6isobar}
\end{figure}

To understand the orientational ordering we can plot the behavior of it's mean value, $\tilde q_6$, for each particle as function of the mean value of $q_4$~\cite{Hernandes22,steinhardt1983,lechner2008,eslami18,geiger2013,boattini2018,boattini2019}. In Fig.~\ref{fig:q6isobar}(a) each point corresponds to one particle in the system in the isobar $p=0.500$ at $T = 0.50$ (black points, BCC phase) and $T= 0.075$ (red squares, fluid phase). As we can see, the sixfold ordering drastically decreases upon cooling. Similar behavior can be seen in the isobar $p=2.00$,(c), where there is a drastic change in the OOP values from the solid HCP phase at $T = 0.100$ to the fluid phase at $T = 0.125$. However, in the fluid-amorphous transitions, (b) and (d), the orientational behavior is practically the same, without an abrupt change in the OOP values.

\begin{figure}[h!]
    \centering
    \includegraphics[width=0.475\textwidth]{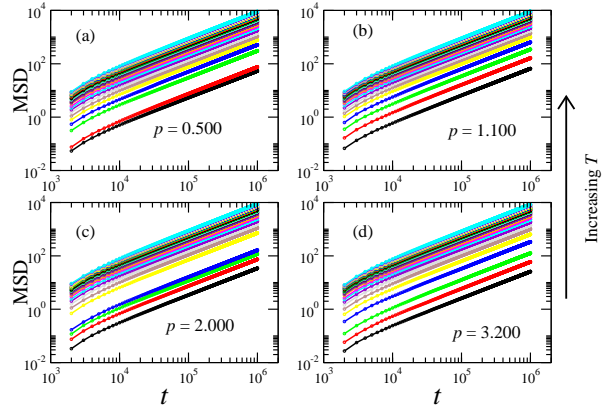}
        \caption{MSD as function of time for pressures that cross (a) the fluid-BCC, (b) fluid-AS1, (c) fluid-HCP and (d) fluid-AS2 transitions. The temperature ranges from $T = 0.025$ up to $T = 0.600$.}
    \label{fig:difftrans}
\end{figure}

Not only the thermodynamic and structural behaviors indicate a discontinuous transition between fluid and crystalline solids and a fluid-amorphous continuous transition, but also the dynamics leads to this conclusion. In Fig.~\ref{fig:difftrans} we show the MSD along distinct isobars for all simulated temperatures. It is clear that the MSD -- and, consequently, the diffusion coefficient $D$ -- have a discontinuity in the fluid-crystalline solid transition, while it has a continuous behavior in the fluid-amorphous solid transition. With this, we can see that the thermodynamic response functions, the translational order parameter, the two-body excess entropy, the OOP and the dynamic of the system indicate the existence of two crystalline solid and two amorphous solid for the potential, Eq.~(\ref{eq:CS}), with discontinuous fluid-crystalline transition and continuous fluid-amorphous transition. Next, we explore the solid-solid transitions in the system.

\subsection{The Solid-Solid transitions}

\begin{figure}[h!]
    \centering
    \includegraphics[width=0.475\textwidth]{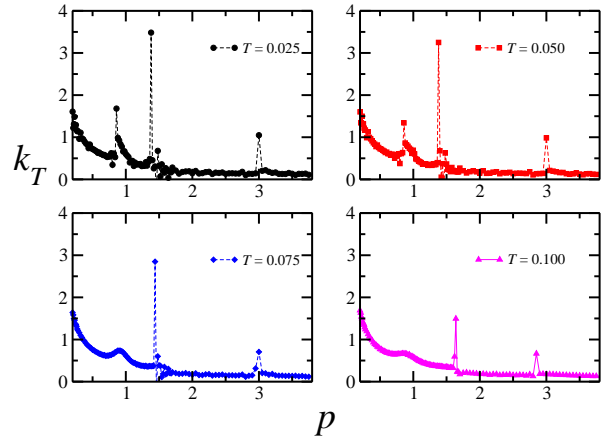}
        \caption{Dependence of $k_T$ as function of $p$ for the lowest simulated temperatures, indicating a continuous transition as the system goes from BCC solid to amorphous solid 1, and discontinuous AS1-HCP and HCP-AS2 transitions.}

    \label{fig:diagramkt}
\end{figure}

 We start this discussion by analyzing the isothermal compressibility, $k_T$, looking for the solid-solid pressure transitions between the four phases: BCC, AS1, HCP and AS2. The BCC and AS1  phases were observed for $T \leq 0.050$, and the HCP and AS2 phases in the $T \leq 0.100$ region. The correspondent $k_T \times p$ curves are shown in Fig.~\ref{fig:diagramkt}. As we can see, at the lowers temperatures, $T = 0.025$ and 0.050, the curve has tree peaks. For $T = 0.025$, they are located at $p = 0.86$, the BCC-AS1 transition, at $p = 1.38$ and $p=3.00$, the AS1-HCP and HCP-AS2 transitions, respectively. Upon heating, at $T = 0.075$ and $T=0.100$, the peak at low temperatures is replaced by a small bump in the $k_T$ curve in the regions near the BCC-AS1 transition observed in the colder isotherms. This bump vanishies at higher temperatures. As shown in the $\rho \times T$ phase diagram at low temperatures, Fig.~\ref{fig:lowTPD}, the extrema $k_T$E line and discontinuous $k_T$D lines correspond to the borders between the crystalline and amorphous solid phases - with the $k_T$E line going to inside the fluid phase, as a prolongation of the BCC-AS1 transition. But how the structural properties are related to this?

\begin{figure}[h!]
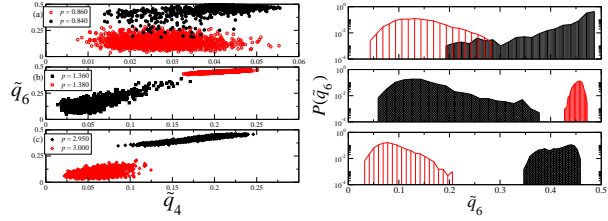

    \centering
     \includegraphics[width=0.2375\textwidth]{diagramaq6q4T0.025.eps}
     \includegraphics[width=0.2375\textwidth]{Q6hist-T0.025.eps}
     
    \caption{Left panel:$\tilde{q}_6\times\tilde{q}_4$ diagram for pressures in the border of the phase transitions along the isotherm $T = 0.025$. Here, each symbol stands for the mean value of each fluid particle. Solid black symbols stand for the lower temperature, and open red symbols for the higher one. From top to bottom, we have the BCC-AS1 transition, AS1-HCP transition and HCP-AS2 transition, respectively. Right panel: correspondent probabilities distribution $\tilde q_6$, $P(\tilde q_6)$. Black curves stands for the lower pressure and red curves for the higher pressure
 }
       \label{fig:q6isotherm}
\end{figure}

First, lets take a look at the orientational ordering, given by the OOPs. In Fig.~\ref{fig:q6isotherm} we show in the left panel the $\tilde{q}_6\times\tilde{q}_4$ diagram for pressures in the border of the phase transitions along the isotherm $T = 0.025$. In the right panel are the correspondent probabilities of a particle have a OOP $\tilde q_6$, $P(\tilde q_6)$. As the system is compressed from $p = 0.840$ to $p = 0.860$ the structure change from a BCC to the AS1, as the $k_T$ curve in Fig.~\ref{fig:diagramkt}. Interestingly, at $p = 0.840$ some particles have an amorphous-like orientational ordering~\cite{lechner2008}, as indicated at Fig.~\ref{fig:q6isotherm}(a). The correspondent $P(\tilde q_6)$, in the right panel, shows a overlap in the probability distributions for both pressures. This indicates that AS1 clusters grow in the BCC phase, promoting the amorphization.  On the other hand, the AS1-HCP and HCP-AS2 transitions have a distinct behavior. As the Fig.~\ref{fig:q6isotherm}(b) and (c) show, there is no overlap in the probabilities, the values of $\tilde q_6$ are distinct for each phase. This indicates a discontinuous orientational transition, with all particles changing from AS1 to HCP and, under compression, from HCP to AS2.

\begin{figure}[h!]
    \centering
    \includegraphics[width=0.475\textwidth]{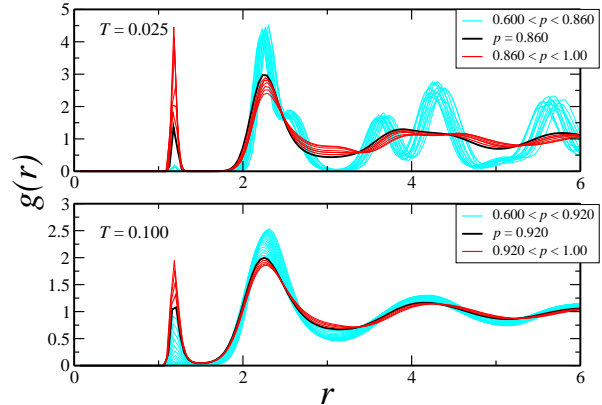}
        \caption{RDF for pressures between 0.600 and 1.00 at temperatures $T = 0.025$ (upper panel) and 0.100 (lower panel). The black curves indicate the RDF where $k_T$ has a maxima, cyan curves are the RDF for lower pressures and in red curves we present the RDF for higher pressures.}
    \label{fig:grtemps}
\end{figure}

From the radial distribution function $g(r)$ is possible to correlate the maxima in $k_T$ at lower pressures with the competition between the two characteristic length scales. The RDF for the isotherm $T = 0.025$, shown in the upper panel of Fig.~\ref{fig:grtemps}, shows how as the BCC phase is compressed the particles start to move from the second length scale, $r_2\approx2.2$, to the first length scale, $r_1 = 1.2$, reaching $g(r_1) \approx 1.0$ at the pressure of transition, $p = 0.860$. This corroborates the idea that an amorphous cluster grows in the BCC phase as the particles start to occupy the first length scale, leading to the observed decrease in $\tilde q_6$ for some particles, and the overlap in $P(\tilde q_6)$ for this transition. At this point, it is relevant to analyze the RDF behavior along the $T = 0.100$ isotherm. At this temperature and in the range of pressures $0.300 < p < 1.00$ the system is in the fluid phase. However, $k_T$ has a maxima  at $p=0.920$ due the small bump. Interestingly, this is the pressure where $g(r_1) \approx 1.0$, following the observed in the BCC-AS1 transition. Although this purely repulsive model does not have fluid-fluid phase transition, it is striking to find this maxima line that converges to the BCC-AS1 transition. Sure, it can not be characterized as the Widom line since it does not end in a LLCP.

\begin{figure}[h!]
    \centering
    \includegraphics[width=0.475\textwidth]{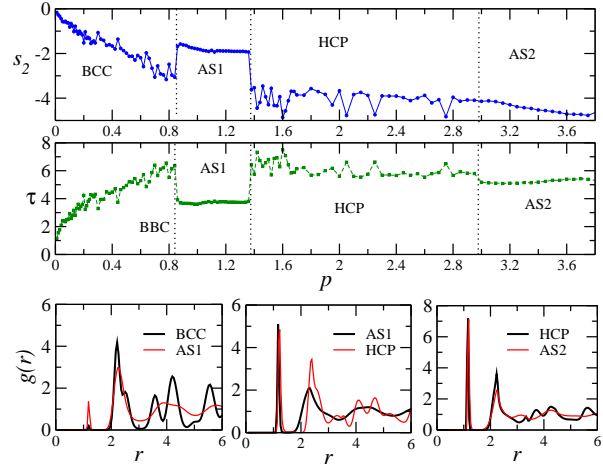}
        \caption{Behavior of $s_2$ (upper panel) and $\tau$ (middle panel) as function of pressure along the isotherm $T = 0.025$. Lower panel: RDF at the phase-transition borders: BCC($p=0.84$)-AS1($p=0.86$), AS1($p=1.36$)-HCP($p=1.38$) and HCP($p=2.95$)-AS2($p=3.00$)}
    \label{fig:s2tauisotherm}
\end{figure}

\begin{figure}[h!]
    \centering
\includegraphics[width=0.475\textwidth]{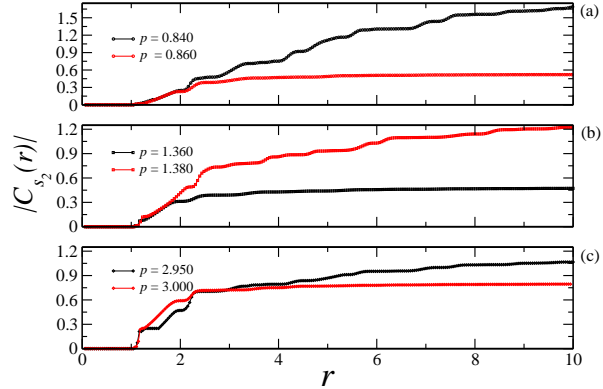}
    \caption{Cumulative two-body entropy in the vicinity of the phase transitions along the isotherm $T = 0.025$. In (a) we show $|C_{s2}(r)|$ BCC-AS1 transition, (b) AS2-HCP transition and (c) HCP-AS2 transition.}
       \label{fig:cums2T0.025}
\end{figure}

Then, as the occupancy in the first length scale grows under compression, the amorphous cluster grows, leading to the order-disorder transition indicated by the $s_2$ and $\tau$ at $T=0.025$ curves, Fig.~\ref{fig:s2tauisotherm} upper and lower panel, respectively. A sharp change in the structure can also be seen at the AS1-HCP transition. However, in the HCP-AS2 phase transition the translational structure varies smoothly -- unlike the OOPs, where the orientation changes abruptly, as the Fig.~\ref{fig:q6isotherm} indicates. Looking to the RDFs at the lower panel of  Fig.~\ref{fig:s2tauisotherm}, we can see that while the BCC-AS1 phase transition is characterized by the particles moving from the first to the second scale, at the AS1-HCP the first scale occupancy did not change, but the ordering for distances from the second scale, $r_2 \approx 2.2$, and beyond are affected. The last solid transition, HCP-AS2, does not have a significant change in the occupancy in the first two coordination shells, but the long range behavior is affected. It can be easily seen in the correspondent cumulative pair entropy, shown in Fig.~\ref{fig:cums2T0.025} for the three phase boundaries.

\subsection{The waterlike anomalies}

\begin{figure}[h!]
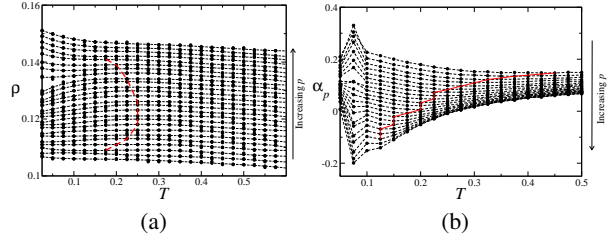

    \centering
     \subfigure[]{\includegraphics[width=0.2375\textwidth]{TMD.eps}}
     \subfigure[]{\includegraphics[width=0.2375\textwidth]{alphapminima.eps}}
    \caption{(a) $\rho \times T$ isobars from $p = 0.500$ (bottom curve) to $p = 1.00$ (top curve). The TMD line is indicated by the red dashed curve. (b) Response function $\alpha_p$ for isobars ranging from $p = 0.300$ to $p = 0.700$. The minima in the fluid phase is indicated by the red line.}
       \label{fig:tanomalies}
\end{figure}

Our findings show that the BCC-AS1 transition is ruled by the competition between the scales, with a sharp change in the translational ordering and the nucleation of a amorphous cluster in the BCC crystal under compression - on the other hand, in the HCP-AS2 transition we have not observed the amorphization of the HCP crystal. The amorphization of the BCC phase seems to affect the fluid phase near the fluid-BCC transition, where maxima in $k_T$ where observed. This indicates that, although this model does not have LLCP, there are two structures, one more BCC-like and another one AS1-like, competing in the fluid phase. Therefore, if there is competition then there should exist waterlike anomalies as well~\cite{alan2008, Stanley2009, gallo2016, Angell14}.

Among the thermodynamic anomalies observed in real water~\cite{url}, here we show the density anomaly, characterized by a maximum in $\rho$ as function of $T$ at constant $p$, and the minimum in $\alpha_p$ - which corresponds to the region where the fluid expand over cooling. The isobars are shown in black in Fig.~\ref{fig:tanomalies}(a) and (b), with the extrema connected by the red curve.

\begin{figure}[h!]
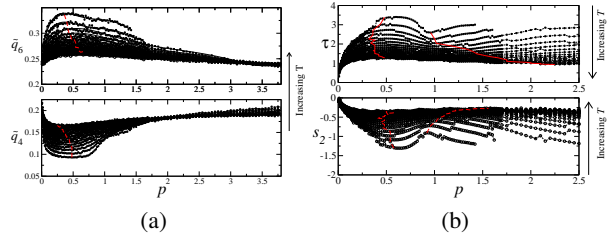

    \centering
    \subfigure[]{\includegraphics[width=0.2375\textwidth]{qanomalies.eps}}
    \subfigure[]{\includegraphics[width=0.2375\textwidth]{diagramatauP.eps}}
    \caption{(a) Orientational and (b) translational order parameter as function of pressure for isotherms in the fluid phase, from $T = 0.025$ up to 0.600. The red lines indicate the extrema in each parameter, delimiting the anomalous region.}
       \label{fig:sanomalies}
\end{figure}

The orientational structural anomaly is usually characterized by a maxima in $\tilde q_6$ under compression, as we observed -- Fig.~\ref{fig:sanomalies}(a). Also, a minima in $\tilde q_4$ was observed in a similar region of the phase diagram. The translational structural anomaly can be observed by an anomalous decrease in $\tau$ under compression - in normal fluids, the translational ordering increases with $p$. Likewise, for regular fluids the pair entropy $s_2$ decreases under compression, but water has an anomalous region delimited by a minima and a maxima in the $s_2\times p$ isotherm. The behavior of both translational parameters are shown in Fig.~\ref{fig:sanomalies}(b).

\begin{figure}[h!]
    \centering
\includegraphics[width=0.475\textwidth]{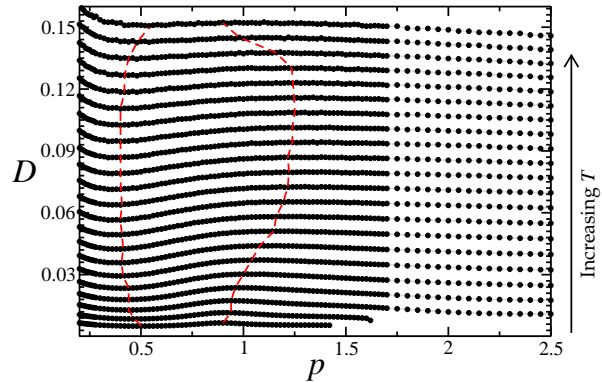}
    \caption{$D\times p$ isotherms from $T = 0.025$ (bottom curve) up to 0.600. The dashed red lines indicate the minima and maxima in $D$.}
       \label{fig:danomalies}
\end{figure}

\begin{figure}[h!]
    \centering
    \includegraphics[width=0.475\textwidth]{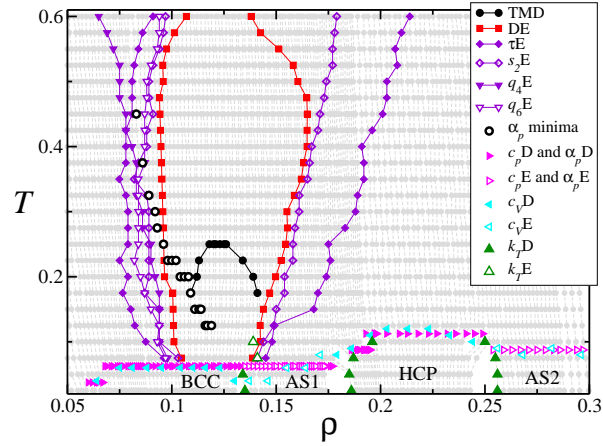}
    \caption{$T\times\rho$ phase diagram. Isobars ranging from $p = 0.050$ up to $p = 4.000$ and $T = 0.025$ to 0.0600 are shown in dashed gray lines with circular points. The Extrema (E) -- that can be a maxima or a minima -- and the Discontinuity (D) observed in the response functions that separate the phases are indicated by the points: magenta filled right triangle for a discontinuous behavior in the $c_p\times T$ and $\alpha_p\times T$ isobars, magenta open right triangle for a extrema in the $c_p\times T$ and $\alpha_p\times T$ isobars, cyan filled left triangle for a discontinuous behavior in the $c_V\times T$ isochores, cyan open left triangle for a maxima in the $c_V\times T$ isochores, green filled up triangle for a discontinuous behavior in the $k_T\times p$ isotherm, green open up triangle for a maxima in the $k_T\times p$ isotherms. The anomalous region are defined by: the TMD line shown by the black filled circles connected by a solid line, the $\alpha_p$ minima indicated by open circles, the diffusion extrema in red squares connected by a solid line, the $\tau$ and $s_2$ extrema lines indicated by filled and open violet diamonds and the $\tilde q_4$ and $\tilde q_6$ extrema by filled and open violet triangles, respectively.}
    \label{fig:fullTPD}
\end{figure}

Finally, the diffusion anomaly is characterized by the increase in the self-diffusion coefficient $D$ under compression. As the $D\times p$ isotherms in Fig.~\ref{fig:danomalies} show, there is a region in the phase diagram where the fluid diffuses faster as the pressure increases. All these anomalies are indicated in the $T \times\rho$ phase diagram, Fig.~\ref{fig:fullTPD}. As in the original works by de Oliveira and co-workers~\cite{DeOliveira2006, BarrosDeOliveira2006}, a waterlike hierarchy of anomalies was obtained as the diffusion, $\tau$, $s_2$, $\tilde q_6$ and $\tilde q_4$ extrema (DE, $\tau$E and $s_2$E, $\tilde q_6$E and $\tilde q_4$E curves, respectively) show. At last, it is interesting to notice how the diffusion and structural anomalous region limit begin at the point were $k_T$ has a maxima, and $g(r_1)=1.0$, reinforcing the idea that the competition between two fluids leads to the anomalous behavior.

\section{Conclusions and Perspectives}
\label{Conclu}

In this paper we have explored the phase diagram of the core-softened potential proposed by de Oliveira and co-authors~\cite{DeOliveira2006, BarrosDeOliveira2006}. We depicted for the first time the solid phases of this fluid, showing that it has solid poly(a)morphism. At low densities, a BCC crystalline phase is observed, followed by an amorphous phase as we compress the system. This ordered-disordered solid transition is a consequence from the competition between the two length scales in the particle's interaction, with an amorphous cluster growing in the BCC crystal , promoting the amorphization. It reflects in the fluid near the BCC and AS1 phases, where we can correlate a maxima in $k_T$ to a transition from a BCC-like fluid  to a AS1-like fluid, where the first coordination shell reaches $g(r) = 1.0$. Compressing the first amorphous phase, the system changes to a HCP crystal phase, with changes in the second coordination shell, and at high pressures, a second amorphous region is observed -- here the long range ordering rules the ordered-disordered transition. However, no amorphous cluster was observed in the HCP phase. Our findings show that even for fluid without LLCP we can find a maxima in the $k_T$ connected to the BCC-AS1 phase transition and to the existence of amorphous clusters in the BCC crystal. It is remarkable that the mechanism observed in the 3D system is similar to what we have observed in the 2D case~\cite{Cardoso21,Nogueira22}. However, in the 2D limit there were more than one anomalous regions, corresponding to ordered-disordered transitions where an amorphous cluster grows inside the crystal. Then, for confined anomalous fluids, there should be a minimal pore size where we will observe a 2D like or a 3D like behavior.

\section{acknowledgments}
Without public funding this research would be impossible. $NpT$ simulations were performed in the Quindim/Camafeu Workstations from the Bordin Lab at UFPel -- proc. 403427/2021-5 from the Brazilian National Council for Scientific and Technological Development (CNpq). JRB is also grateful to the CNPq, proc.  407818/2018-9, and  Research Support Foundation of the State of Rio Grande do Sul (FAPERGS), TO 21/2551-0002024-5, for the funding support. JRB thanks L. C. de Matos for the illuminating insights and L. Pinheiro and T. P. O. Nogueira for the critical reading of the manuscript.   

\section*{Data Availability Statement}

The data that support the findings of this study are available from the corresponding author upon reasonable request.


\end{document}